\documentclass[apjl]{emulateapj}
\shorttitle{XMM observation of V838 Monocerotis}
\shortauthors{Antonini et al.}
\begin{document}
\newcommand\msun{\rm\,M_\odot}
\newcommand\kms{{\rm\,km\,s^{-1}}} 
\newcommand\au{{\rm\,AU}} 
\newcommand\mpc{{\rm\,UPC}} 
\title{XMM-Newton detection of a transient X-ray source in the vicinity  of V838 Monocerotis}
\author{Fabio~Antonini}
\affil{Department of Physics and  Center for Computational Relativity and Gravitation, Rochester Institute of Technology, 85 Lomb Memorial Drive, Rochester, NY 14623, USA }
\author{Rodolfo~Montez~Jr. and Joel~H.~Kastner}
\affil{Chester F. Carlson Center for Imaging Science, Rochester Institute of Technology, 85 Lomb Memorial Drive, Rochester, NY 14623, USA}
\author{Howard~ E.~Bond}
\affil{Space Telescope Science Institute, 3700 San Martin Dr., Baltimore, MD 21218, USA}
\author{Noam~Soker}
\affil{Department of Physics, Technion, Haifa 32000, Israel}
\author{ Romuald~Tylenda}
\affil{Department of Astrophysics, Nicolaus Copernicus Astronomical
  Center, Rabia{n}ska 8, 87-100 Toru{n}, Poland}
\author{Sumner~Starrfield}
\affil{School of Earth and Space Exploration, Arizona State University, Tempe, AZ 85287-1404, USA}
\author{and \\ Ehud Behar}
\affil{Department of Physics, Technion, Haifa 32000, Israel}

\begin{abstract} We report the {\it XMM-Newton}/EPIC detection in 2008
    March of a luminous ($L_X \sim 10^{32-33}{\rm erg}~{\rm s}^{-1}$), variable
    X-ray source in the vicinity (within $\sim6''$) of the enigmatic
    star V838 Mon, which underwent a spectacular outburst in early
    2002. Spectral modeling of the {\it XMM-Newton} X-ray source indicates
    the presence of two plasma components with characteristic
    temperatures of $T_X \sim 2\times10^6$ K and $\sim 1.5\times10^7$
    K, attenuated by an absorbing column ($N_H\sim4\times10^{21}$
    cm$^{-2}$) that is consistent with the visual extinction measured
    toward V838 Mon ($A_V \sim2$). No such luminous source was
    detected in the immediate vicinity of V838 Mon in  {\it Chandra}/ACIS-S
    observations obtained about one year after outburst or, most
    recently, in 2010 January. The two {\it XMM} source spectral components
    appear to be marginally spatially resolved, with the spatial
    centroid of the hard component lying closer to (within $\sim2''$
    of) the position of V838 Mon than the broad-band source or the
    soft source component; 
    however, if there
are two sources at or near V838~Mon, the Chandra nondetections would imply that
{\it both\/} of them are variable. An alternative is that there is a single
variable source, and that the apparent spatial separation may actually be due to
photon-counting statistics or is perhaps instrumental in origin.
     We consider
    constraints placed by the X-ray detection and nondetections on a
    stellar merger model for the 2002 V838 Mon outburst, in which the
    spun-up merger product drives a powerful magnetic
    dynamo. Alternatively, the transient behavior of the X-ray source
    could indicate that the X-rays arose as a consequence of an
    interaction between V838 Mon's ejecta and its early-type (B3 V)
    companion.
\end{abstract}
\subjectheadings{stars: supergiants, stars: novae, stars: individual: V838 Mon, stars: mass-loss}

\section{introduction}
V838 Mon is one of the most enigmatic and unusual objects observed in
stellar astrophysics in recent decades.  It was discovered
undergoing an outburst at the beginning of 2002 January  \citep{BR:02}.
About a month later, its brightness suddenly
increased further, reaching a luminosity of 
 $10^6\rm{L_{\odot}}$ \citep{CR:03} based on a distance of $6~{\rm kpc}$ \citep{SP:08}. 
Although initially believed to be a nova, V838 Mon rapidly
developed distinctly non-nova-like characteristics. In particular, the
spectral evolution clearly showed many peculiar features not
reconcilable with any classical model of stellar eruption
\citep{WA:02,WS:02,RUS:05}.  Similar objects include the luminous red variable M31 RV
in the bulge of the Andromeda galaxy  (Rich et al. 1989; Bond \& Siegel 2006 and references therein), 
and  V4332 Sgr \citep{HA:94,MA:99};
hereafter, we refer to these objects as a single class,
called ``V838 Mon types'' \citep{TY:06}. Only in the case of V838
Mon, however, has the evolution after outburst been well documented
by observations, including {\it Hubble Space Telescope} ({\it HST}) imaging
of its spectacular system of light echoes \citep{BO:03,BO:07}.

After a phase of maximum luminosity, where their spectra resemble 
A-F giants or supergiants, V838 Mon types move on the HR diagram at
roughly constant luminosity and decreasing effective temperature, reaching  spectral type M0, after which the
luminosity rapidly declines.  In the cases of M31 RV, V4332 Sgr and
V838 Mon, an order of magnitude drop in the luminosity was reached
after $\sim 50$, $\sim 10$, and $\sim 100$ days \citep{MO:90, TY:05,
  TY:05a}, respectively.  At the same time, the objects reach their
minimum effective temperature and a maximum photospheric radius of
$\sim 2000\textrm{R}\odot$.  The subsequent evolution proceeds toward
lower luminosity and photospheric radius with a slow increase in
effective temperature.  At the end of this evolution  the V838 Mon types
appear to be late M stars \citep{TY:05,KIM:06,MU:07,KAM:09} or L supergiants  \citep{EV:03}.

Various scenarios have been proposed to explain the
nature of V838 Mon types.
The main contending scenarios are: 
(i) a merger between two main-sequence stars \citep{SO:03};
(ii) a nova-like mechanism involving a thermonuclear runaway on an 
accreting white dwarf \citep{IB:92}; and 
(iii) a He shell flash within a post-asymptotic giant branch (post-AGB)
star \citep{LL:05} . 
\citet{TY:06} discussed these different scenarios, arguing that
the basic characteristics of the V838 Mon eruption are
most consistently explained by  a stellar merger.

The peculiar evolution of V838 Mon types after the observed stellar
outburst provides  evidence against a thermonuclear runaway.  
In  a classical-nova event, the rapidly expanding
ejecta quickly become transparent, exposing a very hot central source; thus the
remnant evolves very quickly to the blue.  By contrast, V838 Mon and similar
objects have evolved to very low effective temperatures and have become
extremely red.

In the He flash mechanism, the runaway process 
occurs in the He-burning shell of an asymptotic giant branch (AGB) or post-AGB star.  Here the
expected typical increase in luminosity is a factor of 2-3 during the
phase of a thermal pulse \citep{WO:81,HE:00} --- much less than the
variation observed in V838 Mon types.

In the case of a stellar merger, if we consider the total
energy released during the V838 Mon eruption, a main-sequence star
would have to accrete a mass of $\sim 0.1$M$_\odot$ on a time scale of
months.  The discovery of an unresolved stellar companion of spectral type B3~V \citep{DE:02,MU:02}
as well as a small cluster of B stars in the vicinity of V838 Mon \citep{AF:07}
 implies that V838 Mon itself has an age $\lesssim 25$ Myr and, furthermore, 
would require that  the system was a triple star, prior to merger.  
Accordingly, \citet{TY:06} showed that the V838 Mon progenitor could
be a $8$M$_\odot$ main-sequence star devouring a $0.3-0.5{\rm M_\odot}$ pre-main-sequence star.

If V838 Mon was initially comprised of a triple stellar system, the
companion would be expected to  play an important role in the dynamics of any stellar
merger and collision.  Resonance interactions (e.g., the Kozai
mechanism Perets and Fabrycky 2009, Antonini et al. 2010 and references therein) between the three bodies could efficiently
shrink the inner binary. Furthermore, the youth of V838 Mon implied by
the presence of a B3-type  companion as well as its membership in a young cluster argues against any
nova-like outburst model, since a white dwarf would not have enough
time to form, accrete material, and then erupt in a nova-like event.

If the V838 Mon outburst was caused by the merger of two stars, then
there is reason to expect that V838 Mon might eventually become a luminous
X-ray source. \citet{SO:07} proposed that the rapidly rotating merger product should
become increasingly magnetically active over a period of a few years
after outburst. This post-merger magnetic activity would be due to a
large convective region in the extended envelope of the merger
product, which is expected to rotate fast, giving rise to an efficient
stellar dynamo. Given the strong empirical connection between stellar
surface magnetic flux and X-ray emission (e.g., Pevtsov et al. 2003), one
would therefore expect V838 Mon to be an X-ray source, under the
stellar merger scenario. Furthermore, the peak X-ray flux from this
magnetically governed, post-merger phase should lag well behind that
of the initial, gravitationally driven optical/IR eruption. On the
other hand, such a ``delayed'' appearance of X-ray emission might also
be consistent with interactions between the expanding matter ejected  from V838 Mon
and its B3~V companion \citep{GORAN, BOND:09}.

To determine whether the V838 Mon outburst resulted in X-ray emission,
we obtained a long-exposure ($\sim 100$Ks) X-ray observation of the
region surrounding V838 Mon with the European Photon imaging Camera
(EPIC) instrument on board the {\it XMM-Newton} observatory. We compared
these {\it XMM} results, obtained $\sim6$ years after the V838 Mon outburst,
with those obtained by the {\it Chandra} X-ray Observatory about one
year after outburst , at which time no bright X-ray source was
apparent at the position of V838 Mon \citep{OR:03},  as well as
  with a very recent (2010 January) {\it Chandra} Director's
  Discretionary Time (DDT) observation obtained subsequent to submission of
  this paper. We discuss the implications of our findings on physical
  models for the optical/IR outburst of V838 Mon.

\section{Data and Analysis}

\begin{figure}
\begin{center}
  \includegraphics[angle=0,width=2.5in]{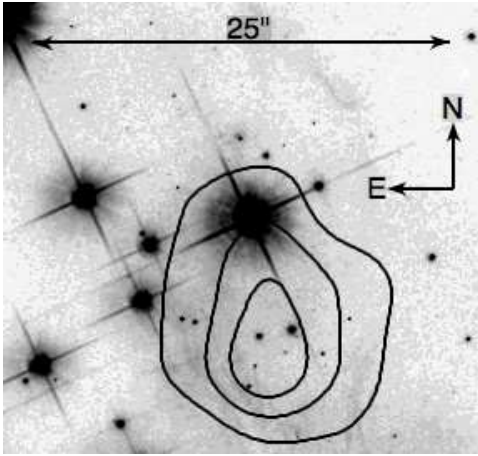}
  \caption{{\it HST} image of the field centered on V838 Mon (the bright star at the center of the field) \citep{SP:08} obtained  in  2005 November 8  (greyscale)  
  in the $V$ filter (F606W) overlaid with contours from the 
  broad-band ($0.5-10$ keV)    XMM-Newton/EPIC 
image of the same field obtained in 2008 March
    (contours at $49$, $66$, and $83$\% of peak intensity.)}
  \label{ImH}
   \end{center}
    \end{figure}

\subsection{2008 {\it XMM-Newton} Observation}

V838 Mon was observed by  {\it XMM-Newton} (ObsID 0500240201) during  {\it XMM}
Revolution 1515, starting on 2008 March 17 at 15:52:37 UT for a total of 110.6, 115.5, and
115.5 ks in prime full window imaging mode with the medium filter on
the EPIC p-type/n-type semiconductor camera (pn) and Metal Oxide
Semi-conductor CCD arrays (MOS1 and MOS2), respectively. We processed
these data with the  {\it XMM-Newton} SAS package, version 7.1.0, and Current
Calibration File Release 241 (XMM-CCF-REL-241). We filtered out
high-background periods and bad events from all observations using
standard filters for imaging mode observations. The resulting net
exposure times are 70.3, 97.8, and 98.2 ks, in the pn, MOS1, and MOS2
arrays, respectively.  

\begin{figure}
\begin{center}
  \includegraphics[angle=0,width=3in]{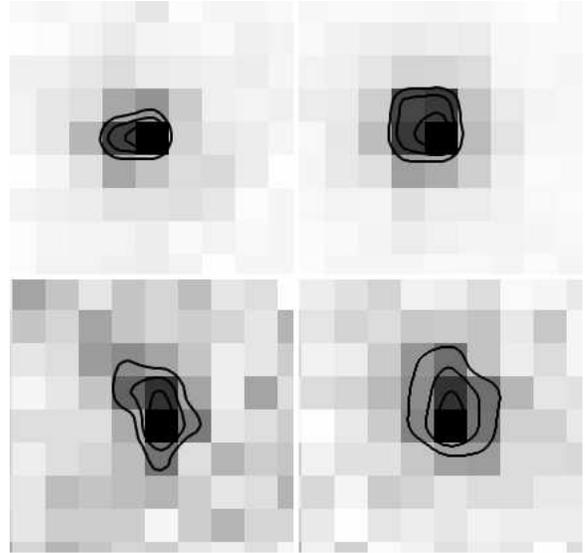}
  \caption{Upper panels: broad band (0.5-10. keV) XMM observation
    (obsID 0551022701) of the point source HD 102195 performed on 2008
    Jun 15. Exposure times were 21.8 ks and 23.4 ks for the MOS1
    (left) and pn (right) detectors, respectively.  Lower panels:
    XMM-Newton/EPIC image of the source overlapping V838 Mon (MOS1,
    left; pn, right).  This source appears somewhat
    asymmetric with respect to the XMM MOS1 and pn PSFs in the two top
    panels.  The images are 42" wide and have north at the top and
    east to the left (contour levels are as in Fig. \ref{ImH}
    ). } \label{comp}
   \end{center}
    \end{figure}

The central region of the resulting, broad-band ($0.5-10$ keV)
 {\it XMM}/EPIC (pn+MOS1+MOS2) image of the V838 Mon field is overlaid on an
{\it HST} image in Fig.~\ref{ImH}. This overlay shows that X-ray
emission is detected very near (within $\sim6''$ of) the position of
V838 Mon, in the  {\it XMM}/EPIC image. Although located nearly on-axis, the
X-ray emission source(s) appears somewhat elongated, such that
although the centroid is displaced from V838 Mon itself, the
lower-level X-ray contours overlap its position.  Fig.~\ref{comp}
compares the morphology of the detected source overlapping V838 Mon
with that of the point source HD 102195 \citep{KAS}.  This star was
observed at a position on the EPIC detectors similar to that of V838
Mon, and its X-ray spectral energy distribution is similar to that of
the V838 Mon source.  The comparison shows that the broad-band source
overlapping V838 Mon's position appears marginally extended and
asymmetric with respect to the  {\it XMM}/pn and  {\it XMM}/MOS1 PSFs (as the MOS2
detector's PSF is somewhat poorer than those of MOS1 and pn, this
image is not shown).   It is worth noting, however, that the
  source extension, although apparent to the eye, is not significantly
  larger than the pn detector's relatively broad PSF ( FWHM $\sim 12.5
  ^{\prime \prime}$ , HEW $\sim 15.2 ^{\prime \prime}$).

To ascertain the accuracy of the absolute astrometry in the  {\it XMM}
observation and the likelihood of a chance association of an X-ray
source with V838 Mon, we performed source detection on the EPIC/pn
image. We correlated the list of X-ray sources detected in the full
(0.5--10 keV) energy range of the  {\it XMM} observation against the
optical/IR USNO NOMAD source list \citep{ZA:04}. Within the ($2800 ~
{\rm arcsec}^2$)  {\it XMM}/pn field there is no apparent systematic shift in
the $\sim100$ correlated NOMAD sources that lie within
$5^{\prime\prime}$ of an X-ray source.  The vast majority
  ($>80$\%) of these NOMAD sources lie within $2^{\prime\prime}$ of
  an  {\it XMM} source, although $\sim$20 sources exhibit larger
  ($\sim5^{\prime\prime}$) displacements.   Among the EPIC/pn
sources, $\sim 50$ sources have $>100$ detected counts ($\sim 5$ times
the photon detection uncertainty of the composite source near V838
Mon), for a source density of $\sim 0.02$ per sq arcmin. Hence, the
probability that the PSF of a random field source would overlap the
position of V838 Mon is $<0.1 \% $.

\subsubsection{Spectral Analysis}

To investigate the nature of the X-ray source overlapping the position
of V838 Mon, we extracted source spectra from the three EPIC detectors
using a circular region with a $25^{\prime\prime}$ radius and centered
on the coordinates of V838 Mon. Background spectra were extracted from
regions near the X-ray emission, located on the same chip of the
relevant detector, avoiding any apparent background sources, and with
a total area on the sky of $\sim 6500^{\prime\prime2}$. We created
target-specific response matrix files (RMFs) and ancillary response
files (ARFs) using the tasks {\it rmfgen} and {\it arfgen} and binned
the spectrum to a minimum of 25 counts per bin. The mean
background-subtracted count rates in the background subtracted spectra
are $7.4$, $1.7$, and $1.4\textrm{ counts ks}^{-1}$ in the pn, MOS1,
and MOS2 observations, respectively. There was variation during the XMM
exposure, however, such that the count rates were a factor $\sim2$
larger than these mean values early in the observation and $\sim30$\%
smaller late in the observation (\S 2.1.3).

\begin{figure}
\begin{center}
  \includegraphics[angle=0,width=3.2in]{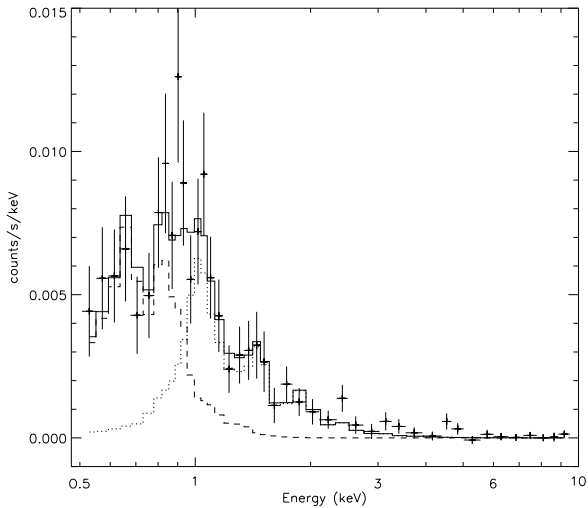}
  \caption{XMM-Newton EPIC pn spectrum of the V838 Mon X-ray source.
    The spectrum (crosses) is overlaid with the result of the best
    simultaneous fit (solid histogram) of a two-component thermal plasma model
    ({\it raymond} model) with the two components subject to different values 
     of intervening absorbing column density ({\it wabs} model).
 The best-fit soft ($\sim2\times10^6$ K) and 
hard ($\sim1.5\times10^7$ K) spectral components are
indicated as dashed and dotted lines, respectively.}
\label{Spec}
\end{center}
\end{figure}

We simultaneously fit the spectra
in XSPEC (ver. 12.3.1; Arnaud 1996) using models consisting of
optically thin, collisional thermal plasma emission ({\it raymond},
Raymond \& Smith 1977) suffering intervening absorption ({\it wabs},
Morrison \& McCammon 1983). The spectrum is reasonably well matched
(reduced $\chi^2 \lesssim 1$) using either single-component or
two-component plasma models. We find, however, that the absorbing
column necessary to achieve acceptable fits using a single-component
model, $N_{H} \lesssim 10^{21}$ cm$^{-2}$, is too small to be
compatible with the optical/IR extinction measured toward V838 Mon
(see below). Hence, given that the XMM source is much more likely to
be associated with V838 Mon than a random (i.e. foreground) field star
(\S 3.1), the single-component model appears inadequate.

  The best-fit two-component model (Fig.~\ref{Spec}) is comprised of a
  ``hot'' plasma in the range $10-16\times10^{6}\textrm{ K}$ and
  a "cooler"  plasma in the range $1.5-2.4\times10^{6}\textrm{ K}$,
  observed through intervening absorbing column densities in the
  ranges $4.9-10\times10^{21}\textrm{ cm}^{-2}$ and
  $3.1-9.2\times10^{21}\textrm{ cm}^{-2}$, respectively (90\%
  confidence ranges). Although the best fit does not tightly constrain
  these resulting values of intervening absorbing column, the 90\%
  confidence ranges of $N_H$ for each {\it wabs} model component are
  consistent with each other and with the range of $A_V$ values toward
  V838 Mon reported in the literature, i.e., $A_V$ $\sim~1.9 ~{\rm
  to}~ 2.7$ \citep{KI:02,MU:05,AF:07}. The observed X-ray fluxes for
the hard and soft spectral components are $9.0\times10^{-15}$ and
$4.2\times10^{-15}\textrm{ erg cm}^{-2}\textrm{ s}^{-1}$,
respectively. The unabsorbed (intrinsic) X-ray fluxes are
$2.9\times10^{-14}$ and $1.8\times10^{-13}\textrm{ erg
  cm}^{-2}\textrm{ s}^{-1}$.  Hence, the hard component has an
  X-ray luminosity of $1.2\times10^{32}(D/6 \rm{kpc})^{2}\textrm{ erg~s}^{-1}$ 
  and the soft component has an X-ray luminosity of
  $7.7\times10^{32}(D/6 \rm{kpc})^{2}\textrm{ erg s}^{-1}$, where the
  adopted distance of $6$ kpc is that inferred for V838 Mon by
  \citet{SP:08}.

\subsubsection{Broad-Band vs.\ Energy-Filtered Source Centroids}

The need for two components to adequately model the EPIC spectra of
the X-ray emission near V838 Mon (Fig.~\ref{Spec}), combined with the
apparent  elongation of this X-ray emission (Fig.~\ref{ImH} and
\ref{comp}), is suggestive of the potential presence of two spatial
components in this source. Since the peak emission is $\sim 1$ keV,
for our spatial analysis we mainly concentrate on the EPIC pn data, as
the pn detector is most sensitive at these energies and has the added
benefit that its PSF, though broader than MOS1 or MOS2, is the most
symmetric of the three EPIC detectors (Fig. 2 \footnote{See also: $
  \rm
  http://xmm2.esac.esa.int/docs/documents/CAL-TN-0029-1-0.ps.gz~~and~~
  \\
  http://xmm.esa.int/external/xmm_{}user_{}support/documentation/$}).
Based on the energy ranges spanned by the two spectral components in
the composite model (Fig.~\ref{Spec}), we created energy filtered
images in the energy ranges 0.5 to 1.3 keV and 1.3 to 10 keV in an
attempt to spatially distinguish the two thermal plasma components.

\begin{figure*}
  \begin{center}
    \includegraphics[angle=0,width=5.2in]{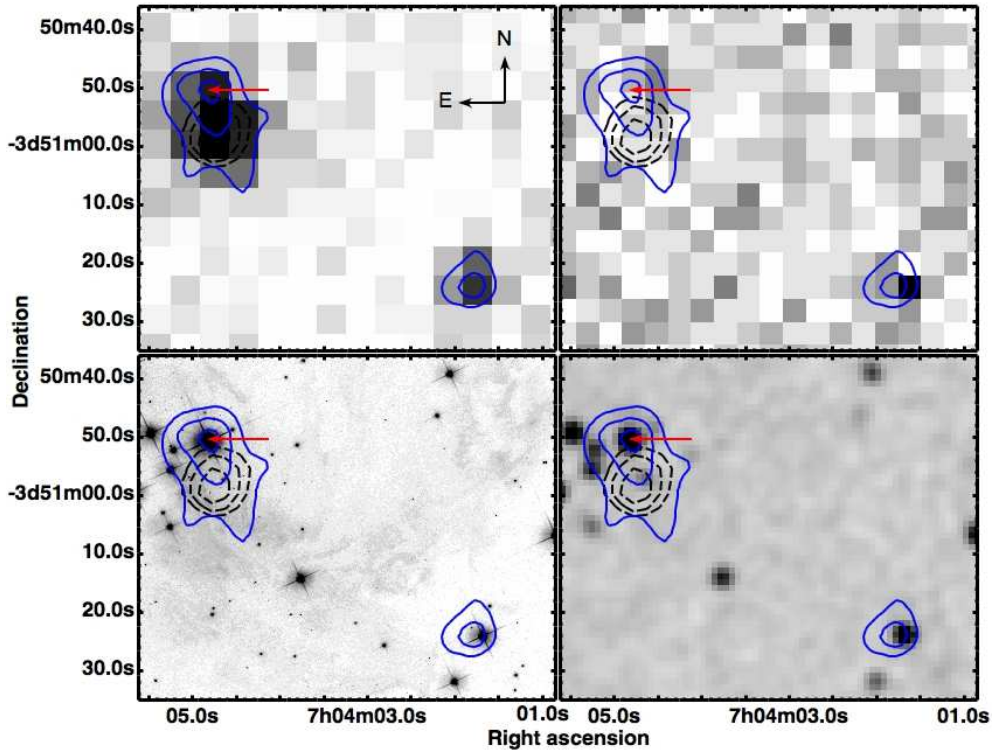}                  
    \caption{Energy-filtered contours of XMM-Newton/EPIC (pn) X-ray
      emission superimposed on greyscale representations of images
      obtained at various optical/IR wavelengths.  The four greyscale
      background images are as follows.  Left upper panel: full energy
      range for the XMM/EPIC (pn) image of the region surrounding V838
      Mon.  Right upper panel: $6.8$ks Chandra image obtained in 2003
      \citep{OR:03}.  Left lower panel: {\it HST} $1$ks exposure
      obtained in 2005 November 8 with the Advanced Camera for Surveys
      in the $V$ filter (F606W).  Right lower panel: 2MASS image in
      the J band.  The arrows indicate the centroid position of V838
      Mon as evaluated from the {\it HST} image.  The smoothed X-ray
      contours were obtained from the XMM-Newton image filtered to the
      energy ranges $0.5-1.3$keV (dashed black contours) and
      $1.3-10$keV (blue contours).  The contour levels are as in
      Fig.~\ref{ImH}.  Unlike the source(s) near V838 Mon, the XMM
      X-ray field source $\approx 55^{\prime \prime}$ southwest of
      V838 Mon was detected during the 2003 Chandra observation (this
      field source was also detected in 2010; see also Fig.\ 7). }
    \label{LF}
  \end{center}
\end{figure*}

\begin{figure}
\begin{center}
  \includegraphics[angle=0,width=3.2in]{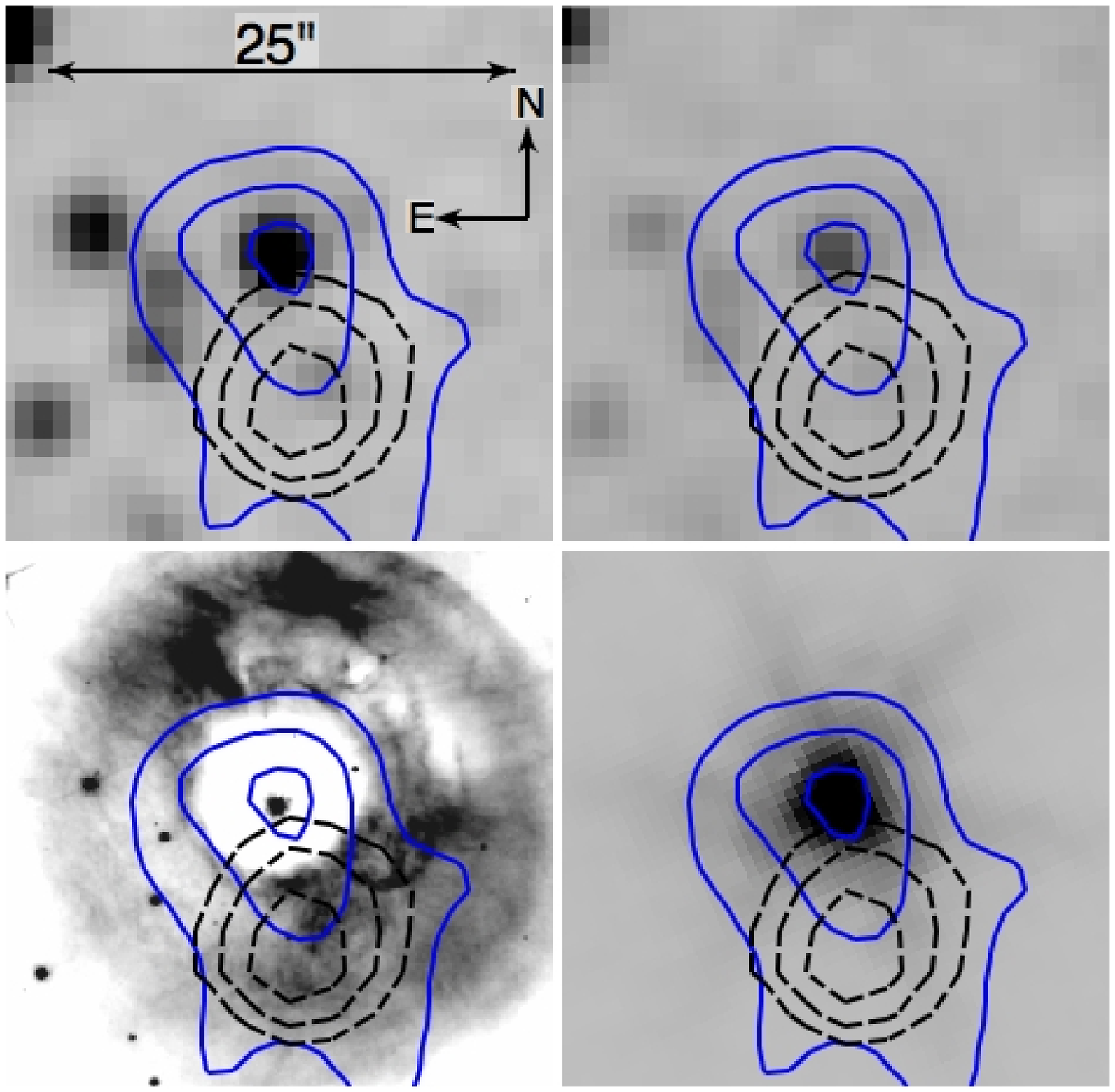} 
   \caption{As for
    Figure \ref{LF}, for the $\sim 30"$ region centered on V838
    Mon. The upper panels are the 2MASS images at J (upper left) and
    $\rm{K_s}$ (upper right). The lower panels show the same region
    but observed with the {\it HST}/ACS in the $V$ filter (F606W) in 2002 May 20 (lower left) 
    \citep{BO:03}  and with
    Spitzer at $3.6\rm{\mu m}$ (lower right) \citep{BA:06}. XMM/EPIC (pn) image contour levels are as in
    Fig. \ref{ImH}. }
\label{SF}
\end{center}
\end{figure}

 We find that these energy filters appear to separate the V838 Mon
  ``source'' into two distinct spatial distributions of X-ray emission
  (Fig.~\ref{LF}, upper left panel); similar energy-filtered images of
  HD102195 do not show such separation.  The overlays of
  energy-specific X-ray emission contours of the soft and hard V838
  Mon sources on {\it HST} and 2MASS images in Fig.~\ref{LF} (lower
  panels) and 2MASS, {\it HST}, and Spitzer images in Fig.~\ref{SF}
  indicate that the hard X-ray flux is coincident with the position of
  V838 Mon to within the $\sim2''$ relative astrometric accuracy of
  the  {\it XMM-Newton} and 2MASS J-band images, whereas the soft X-ray flux
  --- like the elongated, broad-band source itself --- appears to lie
  a few arcsec south of V838 Mon.  If distinct, these two sources
  evidently are not well resolved; their centroids appear to be
  separated by less than the FWHM of the pn PSF.

  In order to compare the position of the broad-band X-ray source with
  that of V838 Mon, as well as to test for the presence of two
  distinct spatial components, source detection was performed
  independently on the V838 Mon X-ray source within broad (0.3--10
  keV), soft (0.3--1.2 keV), and hard (1.5--10 keV) energy bands.
  Table 1 lists the positions of the broad-band, soft, and hard source
  centroids as well as their positional uncertainties and separation
  from V838 Mon. The results indicate that the hard-band emission
  emanates from within $\sim2''$ of the position of V838 Mon, i.e.,
  that the hard-band source is coincident with V838 Mon to within the
  respective (centroiding and astrometric) positional uncertainties.
  The broad-band and soft-band centroids lie $\sim6''$ to the south of
  V838 Mon, however, closer to (within $\sim2''$ of) the position of
  2MASS point source J07040465-0350572 (which also appears as a faint
  star in the {\it HST} images). The displacement of these centroids
  from that of the hard-band source centroid as well as V838 Mon
  itself is well within the FWHM of the pn PSF, but is $\sim3$ times
  larger than the centroiding and astrometric positional
  uncertainties.

 An X-ray source located $\sim 55^{\prime\prime}$ west-southwest
of V838 Mon is coincident with the 2MASS point source
J07040168-0351239 (Fig.~\ref{LF}, upper right panel). This object is
the brightest of a handful of X-ray sources within $\sim1'$ of
  V838 Mon that are detected in the  {\it XMM}/EPIC observation as well as
in both (2003 and 2010)  {\it Chandra}/ACIS-S observations ( see \S
  2.2.2).  The coincidence of  this field X-ray source with the
optical/IR 2MASS point source reinforces the astrometric
registration of the  {\it XMM} imaging.

\subsubsection{X-ray Light Curve}

\begin{table*}\begin{center}
\caption{~~~~~~~V838 Mon: Positions of X-ray Source Centroids}   \label{table}
\begin{tabular}{|ccccc|} 
\hline
\hline
energy range & RA(J2000.0) & DEC(J2000.0) & uncertainty ($^{\prime\prime}$) & offset ($^{\prime\prime}$)  \\
\hline
... &          07:04:04.85$^a$ & -03:50:51.1$^a$      & ...        & ... \\ 
\hline 
0.5 - 1.2 keV	                          &	 07:04:04.71 & -03:50:56.9     &	0.5	&	6.2	\\
1.5 - 10 keV			       &	 07:04:04.79 & -03.50:52.9       &	   1.2	&	2.1	\\
0.5 - 10 keV 			       &          07:04:04.71 & -03:50:56.5     &   0.5   &      5.8 \\
\hline 
a) \citet{BR:02}
\end{tabular}\end{center}
\end{table*}

\begin{figure}
\begin{center}
  \includegraphics[angle=0,width=3.2in]{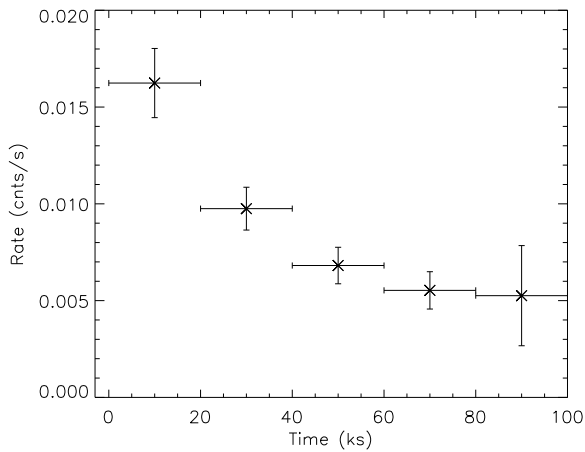} 
  \caption{XMM-Newton EPIC pn light curve of the spectral composite
    source near V838 Mon.  The UT time
    corresponding to t=0 is 15:52:37 on 2008 March 17.  The light
    curve has a bin size of 20 ks and includes X-rays in the energy
    range of 0.5 to 10 keV. A large increase in the background count
    rate occurred toward the end of the $\sim100$ ks exposure,
    resulting in the larger error bar on the last point of the light
    curve. }
\label{LC}
\end{center}
\end{figure}

We extracted source and background X-ray light curves from the
 {\it XMM-Newton} EPIC pn observation, with a temporal bin size of 20 ks. We
present the resulting background-subtracted light curve in
Fig.~\ref{LC}.  The ($25''$ radius) source region used to
  extract the light curve in Fig.~\ref{LC} encompasses the broad-band
  X-ray source overlapping the position of V838 Mon and covers the
  full energy range of pn sensitivity (0.5--10 keV). Ignoring the last
  (highly uncertain) data point, the light curve displays a steady
  decline in X-ray flux from the source near V838 Mon over the
  duration of the observation. The declining X-ray flux may indicate
  that a flare occurred at or just before the beginning of the
  observation, although there is no statistically significant trend in
  the spectral hardness of the source with time. 
   Due to the small number of source photons, we are
also unable to establish whether there was a centroid shift during the
observation,   despite the marginal evidence for a displacement between the soft
  and hard source centroids (Table 1).

\begin{figure*}
\begin{center}
  \includegraphics[angle=0,width=6.5in]{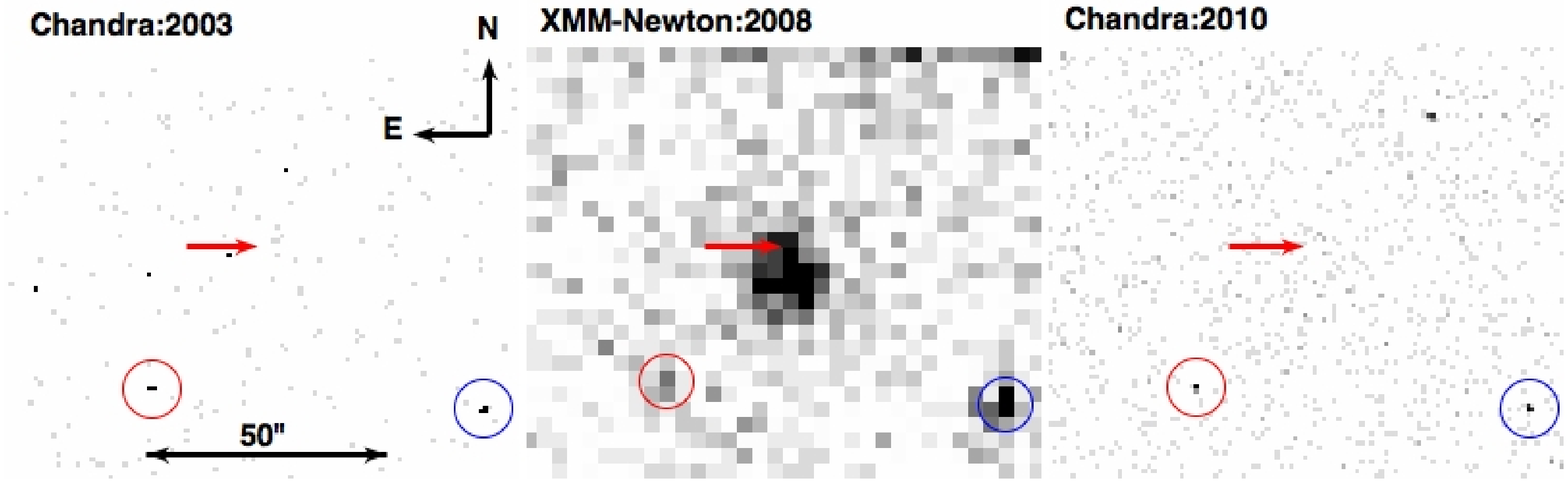} 
  \caption{Left panel: $6.8$ks Chandra/ACIS-S image obtained in
    February 2003.  Middle panel: full energy range for the XMM/EPIC
    (pn) image of the region surrounding V838 Mon obtained in 2008
    March 17.  Right panel: $25$ks Chandra/ACIS-S image obtained in
    2010 January 17.  Red arrows indicate the position of V838 Mon.
    Blue and red circles indicate, respectively, field X-ray
    sources located $\approx 55^{\prime \prime}$ southwest and
    $\approx35^{\prime \prime}$ southeast of V838 Mon
    that were detected in all three observations. }
\label{new}
\end{center}
\end{figure*}

\subsection{2003 and 2010  {\it Chandra} Observations}

\subsubsection{2003 February}

Our reanalysis of the 6.8 ks {\it Chandra}/ACIS-S observation of the
field centered on V838 Mon obtained in Feb. 2003 confirms that no
source was detected within $\sim0.5'$ of V838 Mon.  This observation
imposes a 3$\sigma$ count rate upper limit of $<1.5$ ks$^{-1}$ on any
X-ray source within a $\sim10''$ radius of V838 Mon. Adopting the
best-fit parameters for the hard and soft components deduced from {\it
  XMM}/EPIC spectral fitting (\S 2.1.1), this upper limit translates
to a (3$\sigma$) upper limit of $<7 \times10^{-15}$ erg cm$^{-2}$
s$^{-1}$ for the observed (``absorbed'') flux from each of these
components. Hence, the 2003 {\it Chandra}/ACIS-S exposure likely would
have detected the V838 Mon source if its flux were at the levels found
in the 2008 {\it XMM}/EPIC observation.

  In support of the preceding assertion, we note that the observed
  flux measured for the only source detected by both the (2003)
   {\it Chandra}/ACIS-S and  {\it XMM}/EPIC observations within the field displayed
  in Fig.~\ref{LF} (i.e., the source located $\sim55''$ WSW of the
  composite V838 Mon source; Fig.~\ref{LF}, upper right panel) is
  found to be quite similar in the two observations (as well as in the
  2010  {\it Chandra} exposure described below). Specifically, we
  estimate the 2003 and 2008 fluxes of this source as
  $\sim4\times10^{-15}$ erg cm$^{-2}$ s$^{-1}$ and
  $\sim5\times10^{-15}$ erg cm$^{-2}$ s$^{-1}$ (based on count rates
  of 1.0 ks$^{-1}$ and 2.1 ks$^{-1}$) as measured by  {\it Chandra}/ACIS-S
  and  {\it XMM}/EPIC, respectively.

\subsubsection{2010 January}

After submission of this paper, we obtained a further 24.47 ks DDT
observation of V838 Mon with  {\it Chandra}/ACIS-S on 2010 January 17 (ObsID
12009). The observation was performed in ``faint'' event data mode;
the event data were subject to standard processing, calibration, and
filters (via  {\it Chandra} X-ray Center pipeline v8.2.1 and CALDB v4.1.5,
respectively). The broad-band (0.5--10 keV) image (Fig. \ref{new})
reveals that no X-ray emission was detected within $\sim36''$ of V838
Mon's position.  This observation imposes a 3$\sigma$ ACIS-S count
rate upper limit of $<0.4$ ks$^{-1}$ on any X-ray source(s) at the
position of the source detected by  {\it XMM-Newton} in 2008. Again adopting
the best-fit parameters obtained from spectral fitting of the  {\it XMM}/EPIC
source (\S 2.1.1), we obtain a (3$\sigma$) upper limit of $<2
\times10^{-15}$ erg cm$^{-2}$ s$^{-1}$ for the observed (``absorbed'')
flux from either the hard or soft spectral components.  These limits
correspond to upper limits on the unabsorbed X-ray fluxes of $<3.62
\times10^{-15}$ erg cm$^{-2}$ s$^{-1}$ for the hard component and
$<3.5 \times10^{-14}$ erg cm$^{-2}$ s$^{-1}$ for the soft component,
or  X-ray luminosities of $\lesssim10^{31}$ and $\lesssim
10^{32} {\rm erg ~s^{-1}}$, respectively.  

\section{Discussion}

  The detection by  {\it XMM-Newton} in 2008 of a luminous ($\sim 10^{33}
  {\rm erg ~s^{-1}}$), time-varying X-ray source within a few arcsec of
  the position of V838 Mon (\S 2.1) --- and the lack of detection of
  any source of such high luminosity in  {\it Chandra} observations obtained
  $\sim5$ years before and $\sim2$ years after the  {\it XMM-Newton}
  observation (\S 2.2) --- strongly suggests that the eruption of V838
  Mon left in its wake a luminous, transient X-ray source. In the
  discussion in \S\S 4.2--4.3, we adopt this interpretation. However,
  we first address the possibility that the X-ray source is unrelated
  to V838 Mon and/or is a spatial composite.
  
\subsection{A spatially offset and/or composite source near V838 Mon?}

The spectral/spatial analysis described in \S 2.1.2 indicates that the
broad-band X-ray source centroid exhibits a $\sim6''$ offset from the
position of V838 Mon, somewhat larger than the offsets typical
of correlated optical and X-ray sources in the  {\it XMM} observation (\S
2.1). Hence, the source may be unrelated to V838 Mon. Furthermore, we
find that there may have been two spatially and spectrally distinct
sources of X-ray emission within $\sim6'$ of V838 Mon in
2008. However, we regard both results with suspicion.

First, it seems highly unlikely that any of the other $\sim 4$ random
field stars within the source confusion region could have generated
such strong and variable X-ray emission. As noted previously, the
2MASS source J07040465-0350572 would be most closely associated with
the soft spectral component (if spatially distinct from the hard
component). 
 This 2MASS source could be a reddened, luminous ($\sim6$
$L_\odot$) late-type (K4) member of the V838 Mon cluster \citep{AF:07,
 BO:07a}, on the basis of its visible/IR (HST/GSC2.3/DENIS/2MASS)
magnitudes ($\rm{B}=22.07$, $\rm{V}=20.39$, $\rm{F}=19.26$, $\rm{I}=17.66$,
$\rm{J}=15.910$, $\rm{H}=15.365$ and $\rm{K_s}=15.210$ ) and the visual
extinction to V838 Mon itself (for which we adopt $A_v \sim 2.7$;
\citet{MU:07}). Its association with the soft X-ray component near
V838 Mon would then imply that it is young and highly magnetically
active. However, the X-ray luminosity inferred for the soft source,
$L_x \sim 10^{33}$ erg s$^{-1}$, would be unusually large even for a
flaring pre-main-sequence star, while its X-ray emission temperature
would be unusually low for such an object. If this object is, instead, a
foreground ($D<1$ kpc) star suffering minimal interstellar extinction, it
would be an M dwarf. Such stars do show strong X-ray flares --- although
the flare-to-quiescent flux ratio exhibited by the source near V838 Mon
would be extreme for M flare stars (e.g., Smith, Guedel \& Audard 2005).

More generally, the likelihood that two spatially distinct X-ray sources in
such close proximity to each other could have appeared and disappeared
synchronously is vanishingly small --- as is the likelihood that a random
field dwarf very close to the line of sight to V838 Mon should exhibit
such a large flare.
Furthermore, $\sim20$\% of the
{\it XMM} sources have offsets from their optical/IR counterparts that
are similar to that between the {\it XMM} source near V838 Mon and
V838 Mon itself.

Therefore, given the spectacular optical/IR behavior of V838 Mon since
2002, it seems instead far more logical to conclude that the source
detected very near its position in 2008 by  {\it XMM-Newton} consists of a
single, spectrally complex and highly variable X-ray source associated
with V838 Mon itself.  We note that such an interpretation would
require that the marginal hard-soft PSF offset measured in the
 {\it XMM-Newton} images (\S 2.1.2) is spurious, 
 and is perhaps due to the decreasing signal to noise at higher
energies (70\% of the $\sim610$ hard counts are attributed to the
background while only 30\% of the $\sim470$ soft counts are attributed
to the background).
  Alternatively, such a spurious offset could be instrumental
in origin, although we are unaware of any instrumental effect that
might have produced such a result.

\subsection{The 2008 X-ray source at V838 Monocerotis: constraints on
    outburst models}

  Evidently, the X-ray source at V838 Mon is highly variable, on
  timescales ranging from hours (Fig.~\ref{LC}) to years (Fig.\
  7). Specifically, the  {\it XMM}/EPIC light curve (Fig.~\ref{LC}) provides
  evidence that the source may have been undergoing a long-duration
  ($\stackrel{>}{\sim}40$ ks), energetic flare event, with a peak
  luminosity of $\sim 10^{33} {\rm erg ~s^{-1}}$ and a significant hard
  ($T_X \sim 10^7$ K) component, during the 2008 March  {\it XMM}/EPIC
  exposure. Meanwhile, the nondetection of the source by  {\it Chandra}  in
  2010 January places a firm upper limit of $< 10^{32} {\rm erg~
    s^{-1}}$ on the level of ``quiescent'' emission from this source.
  We now consider these results in the context of the stellar merger
  model for the 2002 V838 Mon optical/IR outburst as proposed by
  \citet{SO:07}.

As described in \citet{SO:07}
the  general relationship between stellar dynamo strength and X-ray
luminosity is
\begin{equation}
L_{X}/L_{bol}=C_{X}R_0^{-2},
\end{equation}
with Rossby number ($R_0$) in the range $0.15 \lesssim R_0 \lesssim
10$ and $C_{X}$ a constant depending on the stellar structure. For G
giants $C_{X} \sim 10^{-6}$ \citep{GO:05} while in the case of main-sequence (MS) stars $C_{X} \sim 10^{-5}$ \citep{PI:03}. Pre-MS stars,
instead, are in or near the saturated regime with typical
$L_{X}/L_{bol} \sim 10^{-3}$ \citep{PR:05}. Adopting conservative
assumptions regarding the level of magnetic activity appropriate for
giant stars, \citet{SO:07} showed that the maximum luminosity of X-ray
emission from V838 Mon would be $L_{X}\sim 3 \times 10^{30}\textrm{erg
  s}^{-1}$ .  If we instead adopt the scaling between $R_0$ and
$L_{X}/L_{bol}$ appropriate for main-sequence stars, the predicted
peak becomes $L_{X} \sim 3\times10^{31}\textrm{ erg s}^{-1}$, while
assuming the (saturation regime) scaling appropriate for pre-MS stars,
the maximum would be between $10^{34}$ and 10$^{35}$ erg s$^{-1}$,
adopting $L_{bol} \sim 2\times10^4$ $L_\odot$ \citep{KAM:09}. 

 The conservative upper limit imposed on the ``quiescent'' level of X-ray
emission by the 2010  {\it Chandra} observation, $\lesssim 10^{32} {\rm erg
  s^{-1}}$, would then imply that the putative V838 Mon merger remnant
is well below the saturated regime, i.e., the remnant is not
pre-MS-like in terms of its magnetic dynamo properties. On the other
hand, the ratio of source $L_X$ in its flare-like state (in 2008)
relative to that in its ``quiescent'' state (in 2003 and 2010),
$L_X$(flare)$/L_X$(quiescent)$>10$, would seem to indicate that the
putative merger has left V838 Mon in an state of extreme magnetic
volatility resembling that of deeply embedded young stellar objects
\citep{Tsujimoto2005}.

The \citet{SO:07} model also predicts that there should be a time lag
between the merger event itself and the generation of a magnetic
dynamo in the merger product (via spin-up of the envelope of the
merger product) that is sufficiently energetic to produce an
X-ray-bright corona, and that the phase of maximum luminosity is
followed by a period of slow decline, with a characteristic time scale
$\sim 100{\rm yr}$. The ``delayed'' appearance of an X-ray source
  at the position of V838 Mon (absent in 2003; present in 2008), taken
  at face value, might appear to offer early support for such a
  model. However, the disappearance of the X-ray source by early 2010
  raises the possibility that the source may have displayed other
  strong but short-lived X-ray outbursts (such as the one detected by
   {\it XMM-Newton}) between the optical/IR outburst in 2002 and the first
  (2003  {\it Chandra}) X-ray observation of V838 Mon. So there is in fact no
  real constraint on the timing of the onset of strong, variable X-ray
  emission from V838 Mon. We conclude that one should be very cautious
  in interpreting the 2008  {\it XMM} source at V838 Mon as supporting (or
  refuting) the "reborn dynamo" model. 

On the other hand the detection of a transient X-ray source by
 {\it XMM-Newton} appears incompatible with He flash models for the V838 Mon
outburst, as AGB stars are not associated with such luminous, variable
X-ray sources (e.g., Kastner \& Soker 2004a) unless such stars reside
in binary systems involving  transfer of AGB mass loss (e.g., 
  Karovska et al.\ 2005).

\subsection{An interaction with the B3 V companion?} 

One possible alternative to the merger scenario as a source of the
X-ray source detected in 2008 is that of the interaction of the
B3~V companion with ejecta from V838 Mon \citep{TY:09}. In December
2006 the companion disappeared from the optical spectrum, only to
reappear again in February 2007 \citep{MU:07b,GORAN}. The eclipse-like
phenomenon followed by the strengthening of the emission line spectrum
can be explained by the occultation of the B3~V companion by a dense
cloud ejected in 2002 and crossing in late 2006 the stellar line of
sight.  The velocities of the expanding ejecta were measured to be in
the range $100-600~ \rm km~s^{-1}$ \citep{CR:03, KI:04, GEB}.
Assuming an outflow velocity of $250~ \rm km~s^{-1}$ \citep{TY:05a},
the time interval between the V838 Mon eruption and the onset of
interactions between its ejecta and its B3 companion star indicates a
binary separation of $\sim 250 \rm AU$.  Later, in September 2007, the
star disappeared again due to its complete immersion into the massive
cloud ejected during the 2002 outburst \citep{BOND:09}.

At the time of the first (2003)  {\it Chandra} observation, therefore,
interactions between V838 Mon ejecta and the B3~V companion had yet to
occur, while by the time of the (2008)  {\it XMM} observation the B3 V
companion was being engulfed inside the expanding matter ejected by V838
Mon. Hence, the late (circa 2008 March) appearance of X-ray
emission from V838 Mon could suggest the onset of ejecta-companion
interactions,  and the subsequent disappearance of the X-ray
  source by 2010 January might then indicate the cessation of such
  interactions.

Establishing the potential role of the B3 V star in the energetic
emission detected by  {\it XMM} in 2008 is beyond the scope of this
paper. However, we note that such high X-ray luminosities cannot arise
from the simple interactions of infalling matter with the matter above
the stellar photosphere unless extreme accretion rates and infall
velocities are involved. The latter would appear to be precluded by
the main-sequence (as opposed to compact object) nature of the
secondary \citep{ShS:88}. Hence other processes --- such as magnetospheric
(reconnection) events generated by the infalling gas or perhaps
wind-wind collisions --- would appear to be required, if the X-rays
were generated via ejecta-companion interactions.  

\section{Summary and Conclusions}

  Our March 2008  {\it XMM-Newton}/EPIC observation of V838 Mon, obtained
  about six years after the spectacular eruption of this enigmatic
  star, reveals a luminous, variable X-ray source centered within
  $\sim6''$ of V838 Mon's position. Spectral analysis of the
   {\it XMM-Newton}/EPIC X-ray source suggests that it consists of a
  relatively hard ($T_X \sim 1.5\times10^7$ K), luminous ($L_X \sim
  10^{32}$ erg s$^{-1}$) component and a softer ($T_X \sim
  2\times10^6$ K) but more intrinsically luminous ($L_X \sim 10^{33}$
  erg s$^{-1}$) component. The absorbing columns inferred
  toward each of these components are similar
  ($N_H\sim4\times10^{21}$ cm$^{-2}$) and are consistent with the
  visual extinction measured toward V838 Mon ($A_V \sim2$). No X-ray
  source(s) of such luminosity was present at or near V838 Mon's
  position during  {\it Chandra}/ACIS-S observations obtained about one year
  after outburst (in 2003) or in 2010. The two  {\it XMM} source spectral
  components appear to be marginally spatially resolved, with the
  spatial centroid of the hard component lying within $\sim2''$ of
  V838 Mon, but the  {\it Chandra} nondetections indicate that the apparent
  spatial separation may be due to photon counting statistics or is
  perhaps instrumental in origin.

The inferred X-ray luminosity  and large temperature range
  inferred for the transient X-ray source at V838 Mon, as well as its
  extreme variability, are consistent with a stellar merger scenario
for the optical/IR outburst of V838 Mon, in which the X-ray emission
arises as a consequence of an energetic magnetic dynamo induced in the
envelope of the merger product. Alternatively, the delayed onset and
subsequent disappearance of the luminous X-ray source near V838 Mon
might be attributed to interactions between matter ejected during the
2002 outburst and V838 Mon's early-type companion. Further X-ray
observations are required to determine the nature of the X-ray
emission detected on March 2008 in the vicinity of V838 Mon.

\bigskip

\acknowledgements{This research has been supported via NASA/GSFC
   {\it XMM-Newton} Guest Observer Facility grant NNX08AD91G to RIT (and
  associated subcontracts to STScI and the University of Arizona). 
 S.S. greatly acknowledges partial support from NASA and NSF grants  
to ASU. 
R.T. acknowledges a financial
support from the grant no. N203 004 32/0448 of the Polish Ministry of
Sciences and Higher Education. The
  authors wish to thank {\it Chandra} X-ray Center Director
Harvey Tananbaum and his Director's Discretionary Time (DDT) team for
their allocation of DDT observing time in 2010 January.
}

\end{document}